\documentstyle[12pt,epsfig]{article}

\newcommand{\la}[1]{\label{#1}}

\newlength{\numlen}

\newlength{\indexlength}

\newcommand{\eq}{\begin{equation}}
\newcommand{\en}{\end{equation}}

\newcommand{\be}{\begin{equation}}
\newcommand{\ee}{\end{equation}}
\newcommand{\ba}{\begin{eqnarray}}
\newcommand{\ea}{\end{eqnarray}}

\newcommand{\fr}[2]{{\frac{#1}{#2}}}

\begin{document}

\begin{titlepage}

\hfill CERN-TH.6918/93\\

\begin{centering}
\vfill

{\bf ON NON-PERTURBATIVE EFFECTS AT THE HIGH-TEMPERATURE ELECTROWEAK PHASE
TRANSITION}
\vspace{1cm}

M. Shaposhnikov\footnote{On leave of absence from the Institute for
Nuclear Research of the Russian Academy of Sciences, Moscow 117312,
Russia.} \\

\vspace{1cm}
{\em Theory Division, CERN,\\ CH-1211 Geneva 23, Switzerland}\\
\vspace{0.3cm}
{\bf Abstract}\\
\vspace{1cm}
\end{centering}
It is argued that confining effects in 3-dimensional non-Abelian gauge theories
(high-temperature limit of 4-dimensional ones) imply the existence of the
condensates of the gauge and Higgs fields in 3-d vacuum. This non-perturbative
effect can decrease the energy of the phase with unbroken symmetry and may
result in the creation of a barrier separating the broken and unbroken phases.
Thus the high-temperature phase transitions in gauge theories can be  stronger
first order than is expected from perturbation theory. The applications of
these results to electroweak baryogenesis are briefly discussed.
\vspace{0.3cm}\noindent

\vfill \vfill
\noindent CERN-TH.6918/93\\
\noindent June 1993
\end{titlepage}

\section{Introduction}
The study of the high-temperature phase transitions in the gauge theories
\cite{kir}--\cite{weinberg} provides an interesting interplay between particle
physics and cosmology. Recently considerable progress has been achieved in the
understanding
of the perturbative expansion for the effective potential of the scalar field
at high temperatures \cite{linde}--\cite{buch}. One can summarize the present
understanding as follows. Let us take for definiteness the standard electroweak
theory with one scalar doublet. Then at zero temperatures the one-loop
effective potential has the form
\eq
V(\phi)_{T=0} = -(2\lambda + B) \sigma^2\phi^2 + \lambda \phi^4
+ B \phi^4 \log (\phi^2/\sigma^2),
\label{vt=0}
\en
where $\phi = \sigma = 250$ GeV corresponds to the minimum $V(\phi)$,
\eq
B=\frac{3}{64\pi^2} (2m_W^4 + m_Z^4 - 4m_t^4).
\en
Here all particle masses are given in terms of the vacuum expectation value of
the Higgs field. The Higgs mass is given by $M_H^2 = 4\sigma^2 (3B+ 2\lambda)$.
For $\lambda \ll g_W^2$ the main contribution to the finite-temperature
effective potential comes from the gauge loops; the
result of the standard calculation in the high-temperature limit ($T$ is much
larger than any mass scale) is \cite{linde,dolan}:
\eq
\Delta V_T = \frac{1}{8} T^2 \phi^2(2m_W^2 + m_Z^2 +2m_t^2) -
\frac{T\phi^3}{4\pi}(2m_W^3 + m_Z^3).
\label{dvt}
\en

The naive loop expansion of the effective potential works provided the mass of
the gauge bosons induced by the Higgs mechanism,  $M_W(\phi) = \frac{1}{2}g_W
\phi$, is larger than the Debye screening scale $M_D = C_D g_W T$ (for the
electroweak theory with three fermionic generations $C_D^2 = \frac{11}{6}$). In
order to go to the smaller values of $\phi$ one has to perform a resummation of
the perturbation theory. This can be done with the use of gap equations for
plasma masses taking into account the Debye screening \cite{eqz,ae,buch}.
Roughly speaking, the effect of the resummation reduces the coefficient in
front of the $\phi^3$ term by a factor $\frac{2}{3}$ \cite{lindeandson} (see
also \cite{ms,carrington}), so that the first-order phase transition is weaker
than one would expect from eq.(\ref{dvt}). The effective potential with this
effect taken into account is
\eq
V_T (\phi) =
\frac{1}{8} (T^2-T_0^2) \phi^2(2m_W^2 + m_Z^2 +2m_t^2) -\frac{2}{3}
\frac{T\phi^3}{4\pi}(2m_W^3 + m_Z^3)+ \lambda \phi^4,
\en
where
\eq
T_0^2 = \frac{8 \sigma^2(B + 2 \lambda)}{2m_W^2 + m_Z^2 +2m_t^2}.
\en
It is argued that for sufficiently small Higgs masses (say, $m_H < 70$ GeV) the
improved perturbative calculation of the effective potential is self-consistent
if $M_W(\phi)> M_M \sim \frac{1}{3\pi}g_W^2 T$ \cite{eqz,ae,buch}. Here $M_M$
is the so-called magnetic mass providing the screening of the static gauge
fields in the plasma, which cures the infrared problem in the thermodynamics of
Yang-Mills fields \cite{linde:pl80,gpy}. So, one expects that the first-order
electroweak phase transition is rather weak for those Higgs masses. For the
larger Higgs masses, the approximation breaks down, although there is an
indication that the phase transition is actually of the second order
\cite{eqz}.

Thus, the perturbative calculations of the effective potential seem to provide
a self-consistent picture of the phase transitions in gauge theories at
sufficiently small scalar self-coupling constants. According to perturbation
theory, the only requirement that has to be met is the smallness of the ratio
$M_M/M_W(\phi)\sim \frac{2 g_W T}{3 \pi \phi}$. This value is indeed small for
the interesting range of temperatures and fields $\phi$ for cosmological phase
transitions in Grand Unified Theories and the electroweak theory, so that one
might think that we achieved (an almost complete) understanding of those
phenomena.

The aim of the present paper is to argue that non-perturbative effects can
change the conclusions drawn from the study of the perturbative expansion of
the effective potential. The paper is organized as follows. In Section 2 we
show that 3-d non-perturbative effects are likely to {\em decrease} the free
energy of the system in the unbroken phase, thus decreasing the value of the
critical temperature. In Section 3 we discuss the contribution of some set of
non-perturbative fluctuations to $V(\phi)$. In Section 4 we analyse the
dynamics of the phase transition in cosmology and argue that the phase
transition can be  actually a very strongly of the first-order, contrary to the
result obtained with perturbative calculation. Some applications of these
results to cosmology and electroweak baryogenesis are discussed in  Section 5.
We summarize our results in the Conclusion.

\section{Non-perturbative effects in non-Abelian gauge theories at high
temperatures}
In what follows we confine ourselves to the study of the phase transition in
SU(2) gauge theory with one Higgs doublet, without fermions. For numerical
presentations we  will use the electroweak gauge coupling, $g_W = 2/3$.  The
results can be easily extended to a more general case.

Our starting point is a dimensionally reduced gauge theory (see
\cite{jackiw}--\cite{landsman} and a recent discussion in \cite{kari}). The
finite-temperature field theory is equivalent (for static quantities such as
the effective potential we are interested in) to a Euclidean field theory, with
compact fourth dimension whose size is $\frac{1}{T}$. Bosonic (fermionic)
fields obey periodic (antiperiodic) boundary conditions in Euclidean time so
that `energies' of particles are discrete ($\omega_B = 2 \pi n T,~~\omega_F =
\pi(2n +1) T,~~n = 0, \pm 1, \pm 2,...$. If all mass scales of the theory are
smaller than the temperature (this is true for the theory with small coupling
constants in unbroken phase and for the system in the broken phase provided the
Higgs-induced mass $M_W$ is small with respect to the temperature) all
fermionic modes and non-zero bosonic modes can be integrated out. In this way
one gets the effective 3-d gauge theory with the action \cite{kari}:
\begin{eqnarray}
&&S_3
= \int d^3x \biggl\{{1\over4} F_{ij}^aF_{ij}^a +
\fr12 (D_iA_0)^a(D_iA_0)^a + (D_i\phi)^\dagger(D_i\phi)+ \nonumber
\\&&
+\fr12 \fr56 g_3^2T A_0^aA_0^a +{g_3^4\over12\pi^2}{17\over16}(A_0^aA_0^a)^2
+
\nonumber
\\&&
+\biggl[-\fr12 m_H^2 +({3\over16}g_3^2 +\fr12\lambda_3)T
\biggr]\phi^\dagger\phi
+\lambda_3(\phi^\dagger\phi)^2 + \nonumber \\&&
+\fr14 g_3^2 A_0^aA_0^a \phi^\dagger\phi \biggr\}+ counterterms,
\la{3daction}
\end{eqnarray}
containing, in addition to 3-d gauge fields and the Higgs doublet, a scalar
field in adjoint representation (former $A_0$ component of the 4-d gauge
field). There are also other higher polynomial contributions to this effective
Lagrangian; they are, however, suppressed by the powers of temperature. In this
equation $g_3^2 = g_W^2 T,~~\lambda_3 = 4 \lambda T$.

The 3-d effective Lagrangian contains four dimensionful parameters. The first
one is just the Debye screening mass $\sim g_W T$; the second scale is a
dimensionful coupling of the 3-d gauge theory $g_3^2 = g_W^2 T$; the third one
is the mass of the Higgs $m(T)^2 = \biggl[-\fr12 m_H^2 +({3\over16}g_3^2
+\fr12\lambda_3)T \biggr]$; and the fourth is the scalar self-coupling. The
phase transition is expected to happen near $m^2(T)=0$, so that at this point
there is a hierarchy of scales, $M_D \gg g_3^2,~M_D \gg m(T)$. Hence, one can
integrate out the $A_0$ field and study the action containing just the 3-d
gauge fields and the Higgs field. From this exercise it is obvious how the
expansion parameter $\frac{g_3}{2\pi M_W(\phi)}$ arises when one computes gauge
loops in the effective potential. Higher-order corrections generate some powers
of $g_3$; one has to compensate these powers in order to get a correct
dimension for the expression, and the only dimensionful quantity at hand is the
mass of the $W$, originated from the Higgs mechanism. As usually in a loop
expansion one gets a number of $2\pi$'s in the denominator from the integration
over momenta.

In some cases, however, the region of applicability of the perturbation theory
{\em cannot} be derived from the analysis of the perturbative expansion. The
well-known example is provided by fermionic number non-conservation in the
electroweak theory \cite{hooft}. Here the amplitudes for the processes with
B-violation are equal to zero in {\em any} order of perturbation theory while
these processes do occur because of instantons. Another example, which is very
close in spirit to the question we are discussing, is provided by QCD. Since we
will use this analogy quite intensively, let us discuss in some detail the
perturbation theory in QCD. As a classic example consider the ratio
$
R = \frac{\sigma(e^+e^- \rightarrow hadrons)}{\sigma(e^+e^- \rightarrow
\mu^+\mu^-)}
$
at some energy scale $s \sim 1$ GeV$^2$ for three fermionic flavours. A
first-order QCD correction to this ratio is known for a long time,
$
R = 2[ 1 + \frac{\alpha_s}{\pi}].
$
The scale at which one should take $\alpha_s$ is not determined at this level.
The computation of the three-loop corrections \cite{kataev} allow one to fix
the scale in $\alpha_s$ and one gets
\eq
R = 2 \left[1 + \frac{\alpha_s(s/\Lambda^2)}{\pi}+ 1.64
{(\frac{\alpha_s(s/\Lambda^2)}{\pi})}^2\right],
\en
where $\Lambda \sim 100-200$ MeV is the scale of the strong interactions in the
$\overline{MS}$ scheme. The convergence of the perturbation theory is pretty
good: at, say, $\sqrt{s} \sim 700$ MeV, one has numerically
$
R = 2[1 + (0.12-0.18) + (0.024 -0.05)]
$
where the first number in brackets refers to $\Lambda = 100$ MeV and the second
one to $\Lambda = 200$ MeV. The natural tendency would then be to conclude that
we know $R$ with  quite a good precision (say, 5\%) at this energy scale. This
conclusion is, however, evidently wrong as can be seen from the experimental
data, since precisely at this scale one has $\rho$-meson resonance and the
correct $R$-ratio has nothing in common with a perturbative expression.

The same story happens with other confining theories, such as  QCD$_2$($N_c$)
in two dimensions \cite{hooftqcd}. Here, at least in the limit of an infinite
number of colours $(N_c \rightarrow \infty)$ the spectrum of the states is
discrete and there is no continuum in the spectrum,  contrary to expectations
from any given order of perturbation theory.

One of the manifestations of confinement is the existence of the
non-perturbative condensates of the different composite operators. For example,
in QCD there are condensates of quark and gluonic fields,
$\langle(\bar{q}q)\rangle \sim -(250 MeV)^3 $ and $\langle \frac{\alpha_s}{\pi}
G^2\rangle \sim (330 MeV)^4$ \cite{vzs}, which appear also in QCD$_2$
\cite{zh}. In QCD in 4-d one can even relate the properties of resonances to
condensates via QCD sum rules \cite{nsvz}; the predictions are quite impressive
for a number of hadronic channels \cite{nsvz}. The analysis of the QCD sum
rules indicates that the actual range of applicability of the perturbation
theory comes not from the analysis of the perturbative expansion but from
non-perturbative corrections.

Now we come back to our 3-d theory. {\em Non-Abelian} gauge theory in three
dimensions has a lot in common with 4-d QCD. Namely, both theories are
confining (one has `logarithmic' confinement in perturbation theory and linear
confinement beyond it in 3d). Therefore, it is natural to assume that the
mechanism for the mass gap generation is the same in both theories \footnote{In
refs. \cite{kks:magn,kks} the $0^{++}$ glueball mass has been estimated in pure
gauge SU(2) theory in 3 dimensions on the lattice. The two papers give
consistent results for the glueball mass; according to \cite{kks:magn} $M_G =
(1.7 \pm 0.4) g_3 $ and according to \cite{kks} $M_G = (2.2 \pm 0.2)  g_3 $.
One can see that, in full analogy with QCD, the scale associated with
non-perturbative effects (3-d glueball mass) is numerically larger than the
perturbative scale $g_3/2\pi$, by a factor $4\pi \sim 10$ (of course, the
parametric dependence is the same). This means that the region of applicability
of the perturbation theory is actually smaller that can be expected from the
analysis of the perturbative expansion.}. In particular, one should expect the
existence of condensates of gauge and scalar fields in 3-d non-Abelian theory.
(The gauge field condensate in a pure 3-d  Yang-Mills theory has been discussed
also in a recent paper by J.M. Cornwall \cite{cornwall} in a different
context.) The lowest-order gauge-invariant operators are $F^2$ (we shall call
it gluonic condensate, in analogy with QCD) and $\phi^{\dagger}\phi$ (scalar
condensate). Just on dimensional grounds, near the point where $m^2(T) = 0$ one
gets
\eq
\langle F_{ij}^a F_{ij}^a\rangle = A_F g_3^6,~~\langle
\phi^{\dagger}\phi\rangle = A_S g_3,~~\langle (\phi^{\dagger}\phi)^2\rangle =
A_4 g_3^2,
\en
where $A_F$, $A_S$ and $A_4$ are some dimensionless numbers. In 4-d QCD, the
numerical values of condensates can be found from the analysis of the
experimental data; in QCD$_2$ from the exact 't Hooft solution \cite{zh}. Here
we do not have this possibility\footnote{One can try to extract the value of
the gluonic condensate from the analysis of QCD sum rules, given the lattice
value for glueball mass in 3d. However, since QCD sum rules do not work in QCD
\cite{nsvz1} and QCD$_2$ \cite{hooftqcd,zh1} for the channel with $0^{++}$
quantum numbers,  we do not attempt to do this here.  Unfortunately, we were
also unable  to extract any information concerning the magnitude of the scalar
condensate from the available lattice data.}.

The knowledge of these condensates allows one to estimate the influence of
non-perturbative effects on the critical temperature of the phase transition.
At critical temperature the free energies of the broken and unbroken phases
coincide; and now we have to take into account the non-perturbative shift of
the energy densities due to the existence of the condensates in the broken and
unbroken phases.  This shift can be easily derived with the scaling arguments.
Indeed, vacuum energy density $\epsilon_{vac}$ in the 3-dimensional theory
given by the functional integral
\eq
\exp(-\epsilon_{vac}V) = \int dA_i(x)\exp(-S_3),
\label{evac}
\en
where $V$ is the 3-d volume and $S_3$ is the action given by eq.
(\ref{3daction}).

Rescaling the fields as $A_i = \sqrt{g_3} \bar{A_i},~~ \phi = \sqrt{g_3}
\bar{\phi}$, and differentiating eq. (\ref{evac}) with respect to $g_3$,
$m(T)^2$ and $\lambda_3$ one gets for small $\lambda_3$ and for $m^2(T)$ close
to zero\footnote{For small $m^2(T)$ and $\lambda_3$ the contribution of the
scalar condensate is small with respect to the contribution of the gluonic
condensate.}:
\eq
\epsilon_{vac} = - \frac{1}{3}\langle \frac{1}{4} F_{ij}^a F_{ij}^a - 2 m^2(T)
\phi^{\dagger}\phi - \lambda_3 (\phi^{\dagger}\phi)^2 \rangle \simeq -
\frac{1}{12}\langle F_{ij}^a F_{ij}^a\rangle,
\label{epsvac}
\en
where we take into account the fact that $\epsilon_{vac} \sim g_3^3$ and used
the equations of motion for the scalar field.
So, again in full analogy with QCD one has an energy shift proportional (with
the minus sign) to gluonic condensate . This energy shift is an analogue of the
bag constant in QCD\footnote{Note that the contribution of the gluonic
condensate to the free energy is of the order of $g_W^6 T^4$, precisely the
term that is not computable by perturbation theory.}. Note that eq.
(\ref{epsvac}) is true in the broken phase minimum as well. The $\phi^3$ term
in the effective potential in this phase arises from the perturbative value of
$\langle F_{ij}^a F_{ij}^a\rangle$.

Clearly, non-perturbative contributions to the energy density should change the
dynamics of the phase transitions at high temperatures. If the gluonic
condensate is positive (as in Euclidean QCD$_4$ \cite{vzs} and QCD$_2$
\cite{zh}), then non-perturbative effects `dig' a `pit' near the origin,
therefore reducing the value of the critical temperature. For the negative
gluonic condensate\footnote{In principle, the gluonic condensate (as well as
the scalar condensate) can have either sign, since the positive-definiteness of
the operator $F^2$ can be spoiled by renormalization.} (this situation seems to
be quite improbable, though) the value of the critical temperature is larger
than that predicted by perturbation theory.

The change of the critical temperature can be estimated by using the value of
the free energy of the system in the broken phase. It is natural to expect that
in the broken phase, with non-zero vacuum expectation value of the Higgs field,
non-trivial fluctuations giving rise to the gluonic condensate are highly
suppressed (we will discuss this point in more detail in the next section), so
that one should take into account this shift in the unbroken phase only. The
critical temperature $T_c$ is  then determined by the equation
\eq
\epsilon_{vac} = V_T(\phi),
\en
where $\frac{dV_T(\phi)}{d\phi}=0$. The shift depends on the mass of the Higgs
boson and constant $A_F$. We present this dependence for positive $A_F$ in Fig.
1. One can see that at $M_H > 50$ GeV and, say, $A_F = 0.1$, the critical
temperature is even lower than the temperature $T_0$, which is thought to be
the temperature of absolute instability of the unbroken phase. Numerically this
shift is quite small for the light Higgs and gets larger for higher Higgs
masses. It is clear that the changing of the effective potential at the origin
generates also a barrier separating the broken and unbroken phases. In
cosmology this barrier will prevent the phase transition to occur at the
critical temperature and it happens at temperatures smaller than $T_c$. In the
next section an attempt is made to estimate this effect.

\section{Estimate of the gluonic condensate}
We are mainly interested  in the non-perturbative contributions to the
effective potential at small $\phi$. It is clear that there is no hope to
determine it in some regular way due to strong coupling and confinement in 3-d
gauge theory. In some models, however, one can study this question by
semi-classical methods in a small coupling regime. The example is provided by
the Georgi-Glashow model in 3-d (SU(2) gauge theory with a triplet of scalar
fields and spontaneous symmetry breaking). It has been shown \cite{polyakov}
that in this model confinement is due to instantons (coinciding with monopoles
in this case). The shift of the vacuum energy density is just
\eq
\epsilon_{vac} =
-\frac{M_W^{\fr72}}{g_3}\alpha(\lambda/g_W^2)\exp\left(-\frac{4\pi
M_W}{g_3^2}\beta(\lambda/g_W^2)\right),
\en
where $\alpha$ and $\beta$ are some functions that can be numerically
calculated, $\beta(0) = 1$, $M_W$ is the $W$-mass. The semi-classical
approximation does not work at small $M_W$, but one can see that this shift is
negative and exponentially small in the broken phase and gets larger with the
decrease of the vacuum expectation value.

The theory with the doublet of the scalar field is more complicated, since
there are no stable solutions to the classical equations of motion even in the
broken phase. So, one cannot perform the analysis in a semi-classical
approximation. To get some insight into the problem, let us study the
non-perturbative gauge field configurations, which may give rise to the gluonic
condensate. We start from a pure Yang-Mills theory. The gluonic condensate is
given by the functional integral
\eq
\langle F_{ij}^a F_{ij}^a\rangle = \int dA_i(x)F_{ij}^a F_{ij}^a\exp(-S).
\label{cond}
\en
According to a suggestion made in \cite{gpy} one can consider the contribution
to eq. (\ref{cond}) of unstable monopole configurations characterized by some
scale $\rho$. To be more specific, we choose as a trial configuration the 4-d
instanton at zero time (see also \cite{rubakov}),
\eq
A_i^a(x) = \frac{1}{g_3}\epsilon_{aij}\frac{x_j}{x^2 + \rho^2}.
\en
The action of this configuration is
\eq
S = \frac{3\pi^2}{g_3^2 \rho}.
\en

There is a number of zero modes associated with this configuration, namely
three translational zero modes and three rotational ones. The normalization
factors associated with them can be computed in  full analogy with the
sphaleron zero-mode problem \cite{armc}. For rotational zero modes we have
\eq
N_{rot} = 8\pi^2 \left[\frac{8 \pi \rho}{3g_3^2}\right]^{\frac{3}{2}}.
\en
Translational zero modes give the contribution
\eq
N_{tr} = \left[\frac{16}{3g_3^2 \rho}(15\pi/8 - \pi^3/12)\right]^{\frac{3}{2}}.
\en
In order to compute the functional integral in the vicinity of this
configuration, one should also introduce a normalization for a scale mode
associated with $\rho$. It is given by
\eq
N_{\rho} = \left[\frac{\pi}{2 g_3^2\rho}\right]^{\frac{1}{2}}.
\en
For dimensional reasons, the determinant of non-zero modes is just
\eq
D = \kappa \frac{1}{\rho^6},
\en
where $\kappa$ is some dimensionless number. From the experience with the
sphaleron \cite{det}, one would expect $\kappa \sim 1$.

Combining all factors together we obtain
\eq
\langle F_{ij}^a F_{ij}^a\rangle =  \int
\frac{d\rho}{\rho}N_{rot}N_{tr}N_{\rho}D\frac{12\pi^2}{g_3^2
\rho}\exp(-\frac{3\pi^2}{g_3 \rho})=0.36 \kappa g_3^6 \sim 0.4 g_3^6 .
\en
The integral over the scale converges at small $\rho$, owing  to an exponential
factor (the energy of the configuration diverges when $\rho \rightarrow
\infty$) and also converges at small $\rho$ due to a pre-exponential entropy
factor. The scale at which the integral is saturated is about $\rho^{-1} \sim 2
g_3^2/\pi^2$. Of course, this estimate of the condensate is rather crude and it
is impossible to compute corrections to it. For our future estimates, we
consider the factor $A_F$ as a parameter of order 1 \footnote{A very different
estimate of the gluonic condensate can be found in \cite{cornwall}, $A_F \sim 5
\cdot 10^{-3}$. Cornwall also concluded that non-perturbative effects introduce
a
negative energy shift in a pure Yang-Mills theory in 3-d.}

Now, we can turn to our theory with Higgs fields. The presence of the scalar
condensate due to the symmetry breaking suppresses the fluctuations of the
gauge field. Again, one can take  't Hooft solution \cite{hooft} for the scalar
field in the vicinity of the instanton with boundary conditions $\phi(x
\rightarrow \infty)= v$
\eq
\phi(x) = \frac{\sigma_i x_i}{(x^2 + \rho^2)^{\frac{1}{2}}}(0,v),
\en
and repeat the previous calculation. In the limit of $m^2(T) = 0$ and of a
small scalar self-coupling constant, one obtains
\eq
\langle F_{ij}^a F_{ij}^a\rangle(\phi) = A_F \frac{g_3^6}{\Gamma({15\over
2})2^{13 \over 2}}z^{15\over 2}K_{15\over 2}(z),
\en
where $K$ is a modified Bessel function and
\eq
z = \frac{3 \pi^2 v}{\sqrt{2}g_3}.
\label{z}
\en
As expected, the non-perturbative contribution decreases exponentially with the
field $\phi$, so that non-perturbative effects are frozen in the broken phase.
Again, it is difficult to estimate the corrections to this calculation;
however, the non-perturbative contribution coming from the gluon condensate has
the properties expected from the physical grounds, namely we are getting a
negative energy shift at the origin, and this shift exponentially decreases in
the broken phase.

To summarize, the effective potential for the scalar field, with the
contribution of the non-perturbative effects and for small $\lambda$ and
$m^2(T)$,  is likely to have an approximate form
\eq
V_{tot}= V_T(\phi) - \frac{T}{12}\langle F_{ij}^a F_{ij}^a\rangle(\phi).
\label{vtot}
\en
This potential is shown in Fig. 2 at different values of the temperature, for
$M_H = 90$ GeV and $A_F = 0.36$, as a function of $z$ and in Fig. 3 for $T=140$
GeV for small values of $z$. With this potential the temperature $T_a$ of the
absolute instability of the unbroken phase is given by
\eq
T_a^2 = \frac{T_0^2}{1 + \frac{16}{13}\pi^2g_W^2 A_F}
\en
and is smaller than $T_0$.

\section{Dynamics of the phase transition}
With  potential (\ref{vtot}) one can study the dynamics of the phase
transition. The crucial question here is the computation of the free energy of
the nucleating bubbles of the new phase. The standard procedure is to find an
extremum of the effective action for the scalar field \footnote{The
non-perturbative effects can also modify the kinetic term for the Higgs field
in addition to the modification of the potential. We do not attempt to estimate
the influence of this effect here.}
\eq
S_{eff} = \int d^3x (\frac{1}{2}\left(\partial_i \phi)^2 + V_{tot}\right)
\en
with boundary conditions $\phi(x) \rightarrow 0, x \rightarrow \infty$, and
$d\phi/dx = 0$ at $x=0$. The phase transition is complete at the temperature
$T_*$ at which the action for the bubble is about \cite{lindenp}
\eq
4 \log\left(\frac{M_{Pl}}{T}\right) \sim 160
\en
for the electroweak theory. As usual, $M_{Pl}$ is the Planck mass.
The problem can be solved numerically. The phase transition occurs at the
temperature $T_*$, smaller than the critical temperature $T_c$. For example, at
$A_F = 1~(0.36,~0.12)$ and $M_H = 130~(90,~60)$ GeV, the  actual temperature of
the phase transition is $T_* \simeq 145~(138,~120)$GeV. The corresponding
values of $T_0$ are equal to $320~(220,~150)$ GeV. The jump of the order
parameter is quite substantial. For all these examples  $\delta\phi \sim 1.6
T$.   So, non-perturbative effects seem to change the behaviour of the
effective potential at the origin and convert a weakly first-order phase
transition to a strongly first-order one, provided $A_F$ is not numerically
small.

Of course, there are many uncertainties in this computation. The main one is
associated with the fact that we are dealing with a confining theory and work
actually in a strong coupling regime. The accuracy of the dimensional reduction
with experimental $g_W$ is quite bad, since there is no hierarchy of scales
$M_M \ll M_D \ll T$, which allow  4-d dynamics to be separated from a 3-d one.
Numerically, all the scales are just the same and $\sim T$. It is not clear how
 this situation should be dealt with. Even inside our approach, at $T=T_*$,
$m^2(T)$ is not small (remember that we neglected it in a previous
computation), and the value of the scalar self-coupling is of the order of
$g_W^2$. We do not think, however that the accounting for the $m^2$ and
$\lambda$ can qualitatively change the conclusion that non-perturbative effects
can make the phase transition strongly first order\footnote{The phase
transition is stronger first order if $\epsilon_{vac}(T_0) <0$ in the unbroken
phase (see eq. (\ref{epsvac})). If $\langle (\phi^{\dagger}\phi)^2\rangle < 0$,
this is always true. If the scalar condensate $\langle
(\phi^{\dagger}\phi)^2\rangle >0$ and large enough, then the transition may be
more weakly first order than in perturbation theory for a sufficiently heavy
Higgs boson, $\lambda_3 > \frac{\langle F^2\rangle}{4\langle
(\phi^{\dagger}\phi)^2\rangle}$.}.

Some support for this observation comes from the 4-d \cite{bunk} and 3-d
\cite{kari} lattice simulation of the electroweak phase transition. In
particular, the data in \cite{kari} indicate that the actual temperature of the
electroweak phase transition is lower than the one predicted by perturbation
theory and that the actual region of metastability is larger. From \cite{bunk}
one sees that the actual jump of the order parameter is also larger than the
one from perturbation theory.

\section{Applications to the electroweak baryogenesis}
All electroweak baryogenesis mechanisms require a first-order phase transition.
Quantatively, the sphaleron mass in the broken phase must be sufficiently large
\cite{s:sm87},
\eq
M_{sph}(T)/T \simeq z_{vac}> 45
\en
in order for the baryonic asymmetry produced at the electroweak transition to
survive to the present time. Here $z_{vac}$ is the expectation value of the
variable $z$ defined by (\ref{z}). This requirement puts an upper bound on the
Higgs mass. The analysis of the one-loop perturbation theory  gives an upper
bound 45 GeV \cite{s:sm87}. As is argued above, non-perturbative effects change
the nature of the phase transition. With the form of the effective potential
advocated in this paper, the critical mass of the Higgs boson may be much
larger and depends on the unknown parameter $A_F$. For example, for $A_F =
1~(0.36,~0.12)$, $M_H^{crit} \simeq 130~(90,~60)$ GeV.  While the uncertainties
in these estimates may be  large, it seems clear that non-perturbative effects
may considerably increase $M_H^{crit}$. In other words, electroweak
baryogenesis can be possible with the experimentally allowed Higgs. At the same
time, the cosmological upper bound on the Higgs mass is still much stronger
than the one originating from triviality arguments. To establish the
cosmological bound with a high accuracy, the better understanding of the
non-perturbative effects in 3-d theories is required. The lattice simulations
can clarify the question. However, as shown in \cite{kari} the lattice size
should be huge in order to take into account all relevant effects.

Another important parameter for the electroweak baryogenesis is the thickness
of the domain wall $a$. With the effective potential derived with the use of
perturbation theory, one gets quite a thick wall. Say, with $M_H = 35$ GeV at
the transition point, $a \sim 40/T$. This value is larger than the mean free
path of the quarks in the plasma, which is estimated to be $l \sim 4/T$
\cite{lindeandson}. With our effective potential, the situation is different.
The domain-wall thickness is roughly the inverse mass of the Higgs boson in the
broken phase at $T=T_*$,
\eq
a \sim \frac{1}{g_w T}\sqrt{\frac{8 {T_*}^2}{3(T_0^2 - T_*^2)}}\sim
\frac{1-3}{T}
\en
for the critical value of the Higgs mass and $A_F = (1-0.1)$. This satisfies
the assumptions, made in the computation of the baryonic asymmetry of the
Universe in the standard electroweak theory \cite{ms,farrar}. To conclude, the
minimal standard model electroweak baryogenesis seems to be in good shape.

The non-perturbative effects in 3-d theory discussed in this paper may change
the dynamics of the grand unified phase transitions. In particular, the amount
of supercooling is expected to be larger than follows from the perturbation
theory. This may result in a prolongation of the inflation period. The study of
these questions is beyond the scope of this paper.

\section{Conclusion}
We have argued that the non-perturbative effects in 3d gauge theories
associated with confinement can change the dynamics of the high temperature
phase transitions in the gauge theories. These effects very likely generate the
condensates of different fields, like gluonic condensate, which decrease the
energy of the vacuum in the unbroken phase. This change results in the decrease
of the critical temperature of the phase transitions and create an additional
barrier separating the broken and unbroken phases. For the standard electroweak
model non-perturbative effects can make the phase transition to be strongly
first order and make an electroweak baryogenesis possible for the
experimentally allowed Higgs boson.

The author is deeply indebted to K. Kajantie, K. Rummukainen and A. Zhitnisky
for many valuable disscussions and suggestions. He is also grateful to K.
Farakos, G. Farrar, A. Kataev and I. Tkachev for helpful comments.

\newpage
\begin{figure}
\vspace{-1cm}
\hspace{2cm}
\epsfig{file=fig1.ps,height=8cm}
\vspace{-3.5cm}
\caption[0]{The dependence of the quantity $\frac{T_c - T_0}{T_0}$ on the mass
of the Higgs boson for different values of the gluonic condensate. Curves 1,2,3
correspond to $A_F = 0,~0.12,~0.36$.}
\end{figure}
\begin{figure}
\vspace{-2cm}
\hspace{2.2cm}
\epsfig{file=fig2.ps,height=8cm}
\vspace{-1cm}
\caption[0]{Effective potential as a function of $z$ for $M_H = 90$ GeV and
$A_F = 0.36$ for different temperatures. Curve 1 - $T=220$ GeV, 2 - $199$ GeV,
3 - $190$ Gev,
4 - $150$ GeV. }
\end{figure}

\begin{figure}
\hspace{1.7cm}
\epsfig{file=fig3.ps,height=8cm}
\vspace{-3cm}
\caption[0]{Effective potential as a function of $z$ for $M_H = 90$ GeV and
$A_F = 0.36$ for $T=140$ GeV at small $z$.}
\end{figure}
\end{document}